# From high-entropy ceramics to compositionally complex ceramics and beyond

Jian Luo [*]

Aiiso Yufeng Li Family Department of Chemical and Nano Engineering; Program in Materials Science and Engineering, University of California San Diego, La Jolla 92093, U.S.A.

## Abstract

Over the past decade, the field of high-entropy ceramics (HECs) has expanded rapidly to encompass a broad range of oxides, borides, silicides, and other ceramic solid solutions. In 2020, we proposed extending HECs to compositionally complex ceramics (CCCs), where non-equimolar compositions and the presence of long- or short-range order, although reducing configurational entropy, create new opportunities to tailor and enhance properties, often surpassing those of higher-entropy counterparts. Along these lines, several fundamental scientific questions arise. *Is the entropy in HECs truly high? Is maximizing entropy always desirable?* In this perspective article, I revisit key concepts and terminologies and highlight emerging directions, including dual-phase CCCs, ultrahigh-entropy phases, and novel processing routes such as ultrafast reactive sintering. I propose that exploring compositional complexity across vast non-equimolar spaces, together with exploiting correlated disorder (coupled chemical and structural short-range order), represents a transformative strategy for designing ceramics with superior performance.

[*] Corresponding Author. Email: jluo@alum.mit.edu (Jian Luo)

# Graphical Abstract

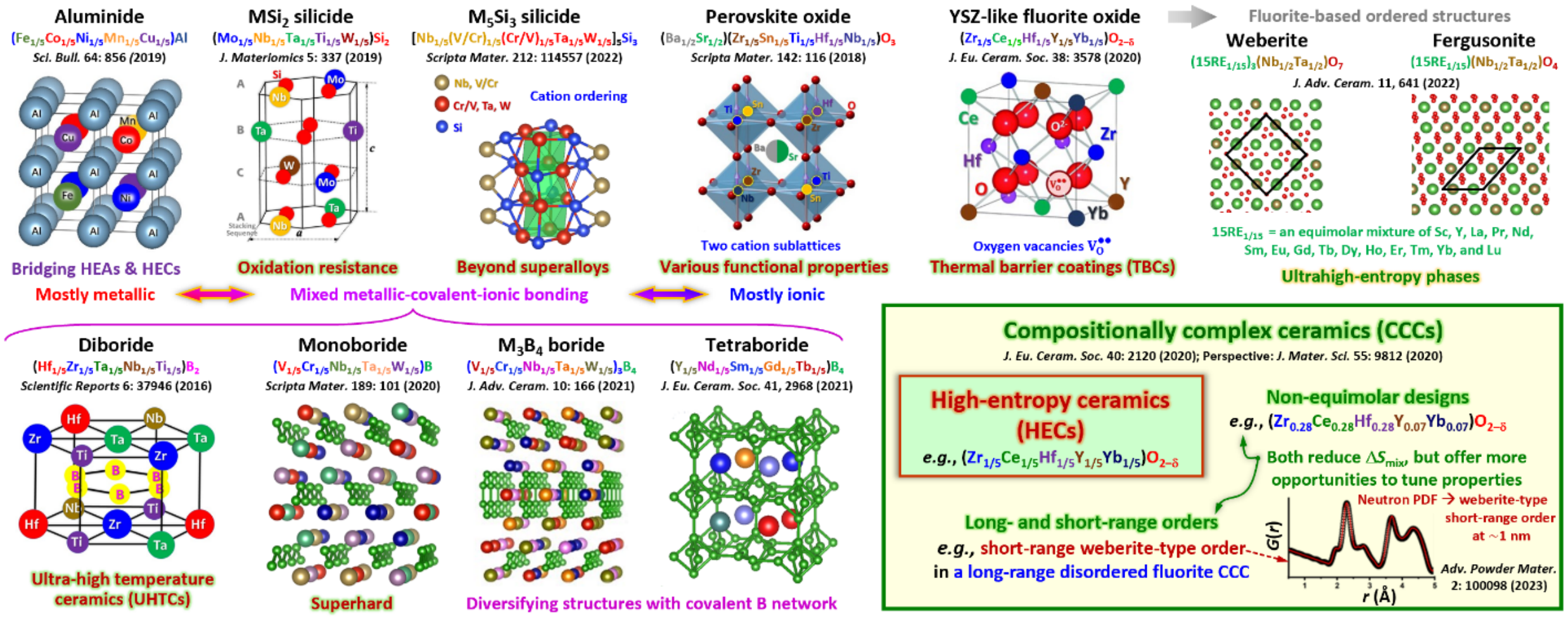

# Highlights

- High-entropy ceramics (HECs) have rapidly expanded across diverse chemistries, crystal structures, and bonding characters.
- HECs extend to compositionally complex ceramics (CCCs), where lower-entropy CCCs can outperform higher-entropy ones.
- Non-equimolar designs offer additional freedom to tune and enhance material properties.
- Short-range order (correlated disorder) in CCCs can enable superior properties.
- Dual-phase CCCs and innovative processing methods, such as ultrafast reactive sintering, open new research directions.

## 1. From High-Entropy Alloys (HEAs) to High-Entropy Ceramics (HECs)

Since the Bronze and Iron Ages, the development of major alloy families, including Cu-, Fe-, Al-, Ti-, and Ni-based alloys, has driven technological and societal advancement. By the early 1900s, however, options for creating entirely new primary-element alloys were largely exhausted. The seminal reports by Yeh *et al.* [1] and Cantor *et al.* [2] in 2004 opened a new paradigm: high-entropy alloys (HEAs), which explore vast compositional spaces with no single dominant constituent [3].

In 2015, Rost *et al.* reported an entropy-stabilized oxide (ESO), $(Co_{1/5}Cu_{1/5}Mg_{1/5}Ni_{1/5}Zn_{1/5})O$, in the rocksalt structure [4], sparking interest in high-entropy ceramics (HECs) as ceramic analogs of HEAs. In 2016, Gild *et al.* reported high-entropy diborides (*e.g.*, $(Ti_{1/5}Zr_{1/5}Hf_{1/5}Nb_{1/5}Ta_{1/5})B_2$) [5] as a new class of ultra-high temperature ceramics (UHTCs), the first high-entropy borides, and the first non-oxide HECs fabricated in bulk form. In 2018, Jiang *et al.* reported the first high-entropy $ABO_3$ perovskite oxides with two cation sublattices [6], stimulating studies on their catalytic, dielectric, ferroelectric, magnetic, thermoelectric, magnetocaloric, and electrocaloric properties, as well as applications in strongly correlated quantum materials, solid oxide fuel cells, protonic electrochemical cells, $H_2O$ splitting, batteries, and supercapacitors. In 2019, Gild *et al.* also reported one of the first high-entropy silicides, $(Mo_{1/5}Nb_{1/5}Ta_{1/5}Ti_{1/5}W_{1/5})Si_2$ [7]. These pioneering studies [4-7] have inspired numerous follow-up investigations worldwide, highlighting enormous opportunities for designing oxide, boride, silicide, and other HECs with novel properties.

Fig. 1 shows representative HEAs [1–3] with body-centered cubic (BCC) and face-centered cubic (FCC) structures, as well as a rocksalt-structured ESO [4], which also adopts an FCC lattice symmetry but features distinct cation and oxygen sublattices. Fig. 1 further illustrates selected families of high-entropy (pervoskite [6] and fluorite [8]) oxides, borides (with $MB_2$ [5], MB [9], $M_3B_4$ [10], and $MB_4$ [11] stoichiometries, where M represents the metallic/cation sublattice, including the first superhard HEC in the MB structure [9]), and silicides (in the $MSi_2$ [7] and $M_3Si_5$ [12] stoichiometries, which may serve as new classes of oxidation-resistant, high-temperature structural materials and coatings), which were first fabricated and reported by our research group. In addition, rocksalt-structured high-entropy carbides such as $(Ti_{1/5}Zr_{1/5}Hf_{1/5}Nb_{1/5}Ta_{1/5})C$, which represent the second class of UHTCs (after high-entropy diborides such as $(Ti_{1/5}Zr_{1/5}Hf_{1/5}Nb_{1/5}Ta_{1/5})B_2$ [5]), were independently reported by multiple research groups in 2018 and 2019 [13–15]. Altogether, these oxides, borides, silicides, carbides, and other HECs exhibit diverse bonding characteristics, crystal structures, and potential applications. Fig. 1 also highlights a single-phase high-entropy intermetallic compound, $(Fe_{1/5}Co_{1/5}Ni_{1/5}Mn_{1/5}Cu_{1/5})Al$, reported in 2019 [16], which serves as a bridge between HEAs and HECs. For further details and references, see an earlier review and perspective article in 2020 [17].

In an extreme extension, recent studies have fabricated and reported 17- to 21-component ultrahigh-entropy fluorite, weberite, pyrochlore, and fergusonite phases, which are illustrated in the bottom-right corner of Fig. 1 and Fig. 3a [18, 19]. For example, Fig. 3a shows a series of 20-component fluorite-based compositionally complex oxides, denoted as "$20CCFBO_{x\text{Nb/Ta}}$", with the general formula $(15RE_{1/15})_{2x+1}(Ce_{1/3}Zr_{1/3}Hf_{1/3})_{3-3x}(Nb_{1/2}Ta_{1/2})_xO_{8-\delta}$ ($0 \leq x \leq 1$), where "$15RE_{1/15}$" represents an equimolar mixture of 15 rare-earth elements (La, Pr, Nd, Sm, Eu, Gd, Tb, Dy, Y, Ho, Er, Tm, Yb, Lu, and Sc) [19]. In this case, although the Gibbs phase rule permits the coexistence of up to 20 phases at thermodynamic equilibrium, this series of $20CCFBO_{x\text{Nb/Ta}}$ compositions all exhibits single ultrahigh-entropy phases with fluorite, pyrochlore, or weberite structures. Notably, this series undergoes an abrupt fluorite–pyrochlore transition at $x \approx 0.27$ and an abrupt pyrochlore–weberite transition at $x \approx 0.87$ with increasing compositional variable $x$, accompanied by discontinuous changes in structural order parameters at both transitions (Fig. 3a) [19].

## 2. From HECs to Compositionally Complex Ceramics (CCCs)

In a 2020 study [20] and a subsequent perspective article [17], we proposed broadening HECs to *compositionally complex ceramics* (*CCCs*) (Fig. 2). In this framework, non-equimolar compositions and the presence of long- or short-range order, both of which reduce configurational entropy, offer additional opportunities to tailor materials properties and, in many cases, to outperform their higher-entropy counterparts [17, 20].

The first study of CCCs [20] investigated the thermal conductivity ($k$) and elastic modulus ($E$) of a series of yttria-stabilized zirconia (YSZ)-like fluorite CCCs, $(Hf_{1/3}Zr_{1/3}Ce_{1/3})_{1-x}(Y_{1/2}Yb_{1/2})_xO_{2-\delta}$. Normally, low thermal conductivity is associated with "soft" materials. However, a distinct and unusual feature of CCCs is the reduction of thermal conductivity accompanied by retained modulus and enhanced hardness, thereby breaking the traditional trade-off between $k$ and $E$. In other words, these materials can achieve low $k$ and high $E$, or an increased $E/k$ ratio, which is not conventionally attainable. For example, the non-equimolar composition $(Hf_{0.28}Zr_{0.28}Ce_{0.28}Y_{0.07}Yb_{0.07})_xO_{2-\delta}$ achieved a higher $E/k$ ratio and demonstrated superior potential as a thermal barrier coating (TBC) material, outperforming its higher-entropy equimolar counterpart $(Hf_{1/5}Zr_{1/5}Ce_{1/5}Y_{1/5}Yb_{1/5})_xO_{2-\delta}$ [20].

A further study revealed that the reduced thermal conductivity of pyrochlore CCCs is governed primarily by a "size disorder" parameter (that represents lattice distortion), rather than configurational entropy, so that medium-entropy compositions can outperform their higher-entropy counterparts [21]. A recent study also suggested that lattice distortion, which is associated with chemical short-range order that reduces configurational entropy, enhances electromagnetic wave absorption in high-entropy diborides [22]. In a broader context, lattice distortion, which does not necessarily scale with configurational entropy, can enhance a wide range of mechanical, thermal, and other functional properties in CCCs. Moreover, weberite-type structural short-range orders can persist within long-range disordered fluorite oxides, enabling ultralow thermal conductivities despite reduced entropy [23]. Taken together, these findings indicate that greater disorder, or maximized entropy, does not always result in enhanced properties.

Extending HECs to CCCs is nontrivial, as it introduces vastly larger compositional design spaces and leverages long- and short-range orders, to tailor and improve materials properties. Furthermore, new physical phenomena can emerge from the interplay among aliovalent doping, anion vacancies, cation ordering, and various types of chemical and structural short-range order (correlated disorder). By exploring compositional complexity and correlated disorder (as elaborated in the next section), CCCs offer even greater opportunities for designing ceramics with exceptional functionalities.

## 3. Correlated Disorder in CCCs

Here, I propose to harness *correlated disorder* [24], particularly coupled chemical and structural short-range order, in CCCs to achieve a range of extraordinary properties. Such correlated disorder can give rise to local symmetries that are distinct from, and typically lower than, the global symmetry [24]. I hypothesize that such correlated disorder can be amplified by the compositional complexity of CCCs, leading to the formation of nanoscale domains with coupled chemical and structural short-range order.

In addition to chemical short-range order (CSRO), which has attracted significant recent interest due to its potential role in influencing the mechanical properties of HEAs [3], we have observed structural short-range order in defect fluorite CCCs. Specifically, a long-range disordered cubic fluorite phase can exhibit weberite-type short-range order or a $\sim\sqrt{2} \times \sim\sqrt{2} \times \sim 2$ superstructure at the ~1 nm scale, as revealed by neutron total scattering [23, 25]. These nanoscale domains (or correlated disorder) can strongly influence ultralow thermal conductivities, and potentially mechanical and other functional properties.

Here, I propose that coupled structural and chemical short-range orders can give rise to correlated disorder in CCCs, which can be harnessed to achieve extreme mechanical, thermal, and functional properties. For example, I hypothesize that correlated disorder (the formation of nanoscale domains) is responsible for:

- Ultralow thermal conductivities in compositionally complex fluorite-based oxides with weberite-type short-range order at ~1 nm [23, 25].
- The unusual hardening effect observed when adding softer, dissimilar $WB_2$ to form $(Zr_{1/5}Hf_{1/5}Ti_{1/5}Ta_{1/5}W_{1/5})B_2$, resulting in enhanced hardness [26].
- The increase in flexural strength of $(Zr_{1/5}Hf_{1/5}Ti_{1/5}Ta_{1/5}Nb_{1/5})B_2$ with increasing temperature, achieving >3× the performance of $ZrB_2$ at 1800–2000 °C [27].
- Ultrahigh dielectric energy storage densities in compositionally complex ferroelectric relaxors, associated with polar nanoregions of reduced size [28].

Further studies are required to elucidate and confirm the mechanisms underlying these extraordinary properties of CCCs and to establish the fundamental principles for harnessing correlated disorder. Such understanding will enable the rational design of materials with superior performance, including (but not limited to) superhardness, exceptional high-temperature mechanical properties, ultralow thermal conductivity, and ultrahigh dielectric energy densities.

As a further example, a recent study characterized chemical short-range order (chemical nanoclusters) in high-entropy borides, which, together with lattice distortion and metal vacancies, enhances electromagnetic wave absorption performance [22].

### 4. Discussion of Terminologies

Analogous to HEAs [3], high-entropy ceramics (HECs) can be defined as ceramic solid solutions with an ideal mixing configurational entropy greater than 1.5 $k_B$ per cation, where $k_B$ is the Boltzmann constant, on at least one cation sublattice (Wyckoff position) if multiple cation sublattices exist. Additionally, medium-entropy ceramics (MECs) can be defined as ceramic solid solutions with ideal mixing configurational entropy in the range of 1–1.5 $k_B$ per cation (on at least one cation sublattice), and low-entropy ceramics (LECs) as ceramic solid solutions with ideal mixing configurational entropy $\leq 1$ $k_B$ per cation (on any cation sublattice). Finally, ultrahigh-entropy ceramics (UECs) are ceramic solid solutions with ideal mixing configurational entropy > 2 $k_B$ per cation on at least one cation sublattice, and UECs are a subset of HECs under these definitions.

We note that these cutoff values (≤ 1 $k_B$ for LECs, 1–1.5 $k_B$ for MECs, > 1.5 $k_B$ for HECs, and > 2 $k_B$ for UECs, per cation) are subjective and somewhat arbitrary. Moreover, these definitions are based on ideal configurational entropies, which do not account for short-range order that can reduce the actual entropy and is often difficult to quantify. It is also noted that equimolar, four-component ceramics are sometimes classified as HECs in literature, even though they are MECs based on the 1.5 $k_B$ per cation cutoff. Therefore, the classifications of LECs, MECs, HECs, and UECs should be considered vague and qualitative, and may not always be critical for practical considerations.

There are multiple ways to define compositionally complex ceramics (CCCs), each referring to a somewhat different scope and therefore meriting discussion. Here, CCCs are first defined as ceramic solid solutions comprising at least three principal components, with an emphasis on exploiting non-equimolar compositions and correlated disorder to tune and enhance material properties, potentially surpassing those of their higher-entropy counterparts. In the field of HEAs, a principal element is typically defined as having an atomic fraction between 5% and 35% [3]. However, this range is somewhat arbitrary (*e.g.*, a 40%–30%–30% ceramic solid solution would not qualify as a CCC under this definition). Here, I propose to alternatively define a principal component as one whose molar fraction is no less than 1/3 of that of the most concentrated component in the solid solution. In this Perspective article, I also propose a second (simpler) definition of CCCs: ceramic solid solutions in which no single component equals or exceeds 50 mol % (*i.e.*, there is no dominant component). Under these definitions, which are inherently subjective, HECs are largely (but not strictly) a subset of CCCs, whereas CCCs are not defined by a specific entropy range. In an entropy-based framework, CCCs may be defined as ceramic solid solutions with an ideal configurational mixing entropy exceeding 1 $k_B$ per cation on at least one cation sublattice. Under this third definition, CCCs constitute a superset of both MECs and HECs (with UHCs as a subset of HECs). Across all three definitions, which are not fully equivalent, the term CCCs highlights opportunities to tailor and enhance material properties through rational non-equimolar compositional designs and long- and short-range order, despite the accompanying reduction in entropy.

If an enthalpic barrier is overcome by an entropic gain to form a phase, this phase can be considered "entropy stabilized". Entropy-stabilized phases can exist in not only HECs, but also MECs or even LECs, and they may be stabilized by different types of entropy. In general thermodynamic terms, a temperature-induced phase transformation at a constant pressure produces a high-temperature phase that is entropy-stabilized, while the low-temperature phase is enthalpy-stabilized. The concept of entropy stabilization in oxides was likely first introduced by Navrotsky in the late 1960s during the thermodynamic analysis of binary $A_3O_4$–$B_3O_4$ spinel solid solutions (that are LECs) [29]. Rost *et al.* proposed a specific and more restrictive set of criteria for defining entropy-stabilized oxides (ESOs) [4]. It should be noted that many high-entropy oxides are not ESOs, and not all ESOs are HECs.

Critical questions often arise regarding HECs: *Is the entropy in HECs truly high? Is maximizing entropy always desirable?* For a five-component equimolar HEC, the ideal configurational entropy is $\ln 5 \cdot k_B$, or ~1.61 $k_B$ per cation (on one Wyckoff position). Its contribution to Gibbs free energy is comparable to thermal vibrational energy (1.5 $k_B T$ per atom from kinetic energy, or 3 $k_B T$ per atom when including both kinetic and potential energies, according to the equipartition theorem). This magnitude may, in some cases, be sufficient to stabilize a single-phase HEC against phase separation or precipitation, as elegantly demonstrated by the case of ESO $(Co_{1/5}Cu_{1/5}Mg_{1/5}Ni_{1/5}Zn_{1/5})O$ [4]. However, the contribution of configurational entropy to the total free energy is generally modest, particularly when compared with high-enthalpy reactions such as oxidation for non-oxide HECs.

## 5. Dual-Phase CCCs

In 2020, Qin *et al.* reported the first dual-phase HECs, consisting of five-component diboride and carbide phases in thermodynamic equilibrium (Fig. 3b) [30]. Interestingly, although each specimen was synthesized with equimolar amounts of Ti, Zr, Hf, Ta, and Nb, equilibrium partitioning between the two phases resulted in non-equimolar cation compositions within both the boride and carbide phases (Fig. 3b). Consequently, each individual phase is no longer "high entropy", and the system is more appropriately described as dual-phase CCCs. A thermodynamic relationship governing the equilibrium compositions of these carbide and boride CCC phases was established, and a corresponding thermodynamic model was developed [30]. This series of dual-phase compositionally complex UHTCs exhibit hardness values exceeding the weighted linear average of the two single-phase high-entropy diboride and carbide counterparts, both of which are already harder than the rule-of-mixtures predictions based on individual binary constituents. This finding highlights the potential of microstructural engineering in dual-phase CCCs to further enhance materials properties.

The introduction of dual-phase HECs/CCCs extends the state of the art, providing a new platform to tailor functional and mechanical properties through controlled variations in phase fractions, compositions, and microstructures. In general, dual-phase CCCs represent a promising new direction in the design of CCCs, offering expanded opportunities for microstructural engineering. It is important to note that the individual CCC phases in dual-phase systems are intrinsically non-equimolar and often not "high-entropy" in five-component systems. This underscores the need for further fundamental studies and thermodynamic modeling to understand and predict dual-phase equilibria in CCCs, ultimately enabling rational design of dual-phase CCCs with tailored microstructures and superior properties.

## 6. Novel Processing: Ultrafast Reactive Sintering as an Example

Another emerging research direction involves novel synthesis and processing methods for CCCs, which are generally more challenging to densify and homogenize due to their compositional complexity. Nevertheless, recent studies have demonstrated intriguing and promising opportunities for ultrafast reactive sintering of CCCs.

As illustrated in Fig. 3c, a single-phase $(Hf_{1/5}Zr_{1/5}Ta_{1/5}Nb_{1/5}Ti_{1/5})B_2$, which represents one of the most difficult ceramics to sinter, was successfully synthesized from a mixture of five binary boride powders and densified to >99% of the theoretical density, yielding a homogeneous single-phase microstructure within ~2 min using a flash spark plasma sintering approach [31]. High-entropy borides have also been synthesized and sintered by ultrafast high-temperature sintering [32, 33]. Since then, an increasing number of high-entropy oxides, carbides, and borides have been synthesized using various reactive ultrafast sintering techniques. These studies demonstrate that homogeneous single-phase ceramics can be synthesized and densified within seconds to minutes—a surprising for compositionally complex refractory systems considered difficult to sinter. Ultrafast reactive sintering was also employed to fabricate porous high-entropy diborides with low thermal conductivity by suppressing coarsening [34].

This emerging capability of ultrafast reactive sintering of CCCs opens new avenues for high-throughput materials discovery and for cost- and energy-efficient fabrication. It also presents new opportunities to produce non-equilibrium phases and unique microstructures via far-from-equilibrium processing. At the same time, it underscores the need for fundamental studies to elucidate the underlying mechanisms and kinetic pathways governing reactive ultrafast sintering in systems with extreme compositional complexity.

## 7. Outlook

Over the past decade, the field of HECs has expanded rapidly to encompass oxides, borides, silicides, carbides, nitrides, fluorides, silicates, and other ceramic solid solutions featuring diverse bonding characters, stoichiometries, crystal structures, and defect chemistries. Extending the concept of HECs to CCCs, which explore vast non-equimolar compositional spaces and exploit correlated disorder, provides new opportunities to tailor and enhance properties, even as entropy decreases.

These rapid and exciting developments naturally raise critical questions: *What superior properties can be achieved in HECs and CCCs? Through what mechanisms?* As the functional properties of HECs span enormous range, a comprehensive review lies beyond the scope of this perspective article. Instead, I offer a few general observations and opinions.

In general, increasing compositional complexity in HECs and CCCs enables the enhancement of several key properties: reduced thermal conductivity and improved thermoelectric performance (*via* enhanced phonon scattering), increased hardness and superior high-temperature mechanical behavior (through impeded dislocation motion), and enhanced catalytic activity (*via* defect engineering).

Beyond these, HECs and CCCs display a broad spectrum of promising functional properties, the underlying mechanisms of which are often complex and not yet fully elucidated. Here, an advantage of HECs and CCCs may lie in providing vast, previously unexplored compositional spaces for tuning and enhancing properties, potentially enabling the optimization of multiple targeted properties (including desired properties that are counter-correlated) that are often difficult to achieve simultaneously in simpler systems. In this regard, compositional complexity, which encompasses non-equimolar designs, correlated disorder, order–disorder transitions, and diverse defect chemistries (*e.g.*, aliovalent doping and vacancies), provides a more powerful and versatile framework for materials design than the mere maximization of configurational entropy. Accordingly, we favor the terminology compositionally complex ceramics (CCCs) over "high-entropy ceramics" (HECs, which is largely a subset of CCCs), as entropy itself is not necessarily the dominant factor controlling property optimization.

Realizing the full potential of CCCs requires predictive models and effective design strategies, including data-driven approaches enabled by high-throughput synthesis, such as ultrafast reactive sintering. Furthermore, microstructural and defect engineering, incorporating dual-phase CCCs and far-from-equilibrium processing routes, will also play an essential role in developing CCCs with unprecedented properties.

Although microstructures and interfaces are key determinants of material properties, their roles in CCCs remain largely unexplored. A transformative pathway for advancing CCCs may be found in exploiting interfacial phase (complexion) transitions and constructing grain boundary (GB) “phase” diagrams [35]. Notably, a recent study revealed a GB disordering transition in compositionally complex perovskite solid electrolytes, which were leveraged to facilitate grain growth and improve ionic conductivity [36, 37]. Along this direction, exploring two-dimensional (2D) interfacial phases (complexions) in CCCs offers a promising route to tailor microstructural evolution and enhance material properties. Moreover, the concept of high-entropy grain boundaries (HEGBs) [38], recently introduced to stabilize nanocrystalline alloys at high temperatures, may be extended to ceramics to unlock further opportunities for interfacial engineering. As a highly encouraging example, analogous high-entropy interfaces may help stabilize porous spinel CCC (Co, Mg, Mn, Ni, Zn)(Al, Co, Cr, Fe, $Mn)_2O_4$, achieving ultralow thermal conductivity and diffusivity (approximately 1000 times lower thermal diffusivity than that of air) while maintaining a high elastic modulus and remarkable microstructural stability at 1000 °C [39].

## 8. Concluding Remarks

Since 2015, the rapid development of high-entropy ceramics (HECs) encompassing diverse chemistries, bonding characters, crystal structures, and defect types has revealed vast opportunities for designing ceramics with novel and enhanced properties. In 2020, we further proposed extending HECs to compositionally complex ceramics (CCCs) to unlock new pathways for property optimization. Here, I propose that exploring compositional complexity across broad non-equimolar spaces, in combination with correlated disorder, offers a promising strategy for designing high-performance CCCs. Furthermore, additional opportunities arise from dual-phase CCCs that enable microstructural engineering, from novel synthesis and processing methods such as ultrafast reactive sintering, and from the exploitation of 2D interfacial phases (complexions) in CCCs to control microstructural evolution and improve material properties.

## Declaration of Competing Interest

The authors declares that he has no known competing financial interests or personal relationships that could have appeared to influence the work reported in this paper.

## Acknowledgements

The author acknowledges current support from the Synthesis and Processing Science Program of the U.S. Department of Energy (DOE), Office of Science, Basic Energy Sciences (BES), Division of Materials Science and Engineering, under Grant No. DE-SC0025255, for research on the ultrafast reactive sintering of compositionally complex ceramics, a key topic discussed in this perspective article. The author is also grateful to current and former students and collaborators for their contributions, as well as to prior sources of support for studies on the fabrication and investigation of various classes of HECs and CCCs over the past decade.

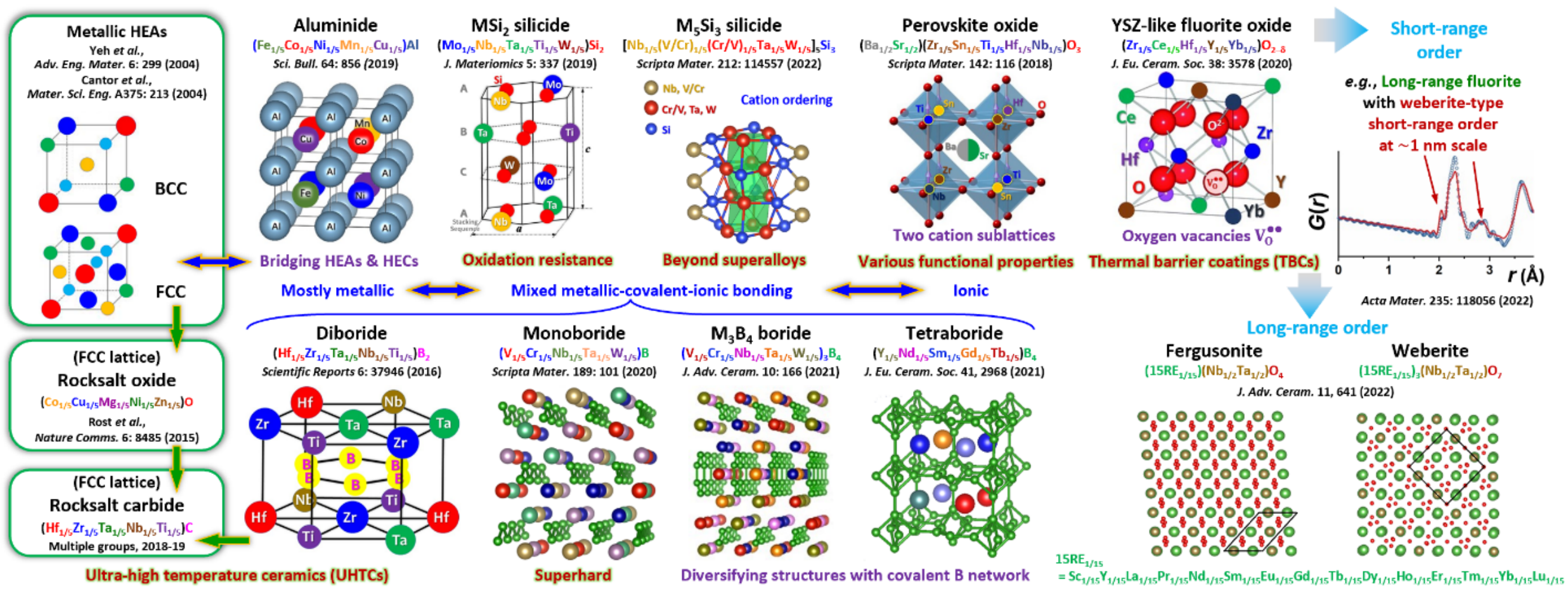

**Fig. 1. The rapid expansion of the field of high-entropy ceramics (HECs).** Shown are representative metallic HEAs [1–3] with BCC and FCC structures, a rocksalt entropy-stabilized oxide (ESO) [4], and a rocksalt high-entropy carbide [13–15], all adopting the FCC lattice symmetry but featuring distinct metal and oxygen/carbon sublattices. Also illustrated are selected families of more complex high-entropy oxides, including $ABO_3$ perovskites with two cation sublattices [6] and YSZ-like fluorite-based structures [8] exhibiting long- and short-range structural orders [18, 23, 25]; borides ($MB_2$ [5], MB [9], $M_3B_4$ [10], $MB_4$ [11], where M represents the metal/cation sublattice); silicides ($MSi_2$ [7] and $M_3Si_5$ [12]); and aluminides [16], which are single high-entropy intermetallic compounds that serve as a bridge between HEAs and HECs [16], first reported by our research group [17]. Schematics were adapted from Zhou *et al.*, Sci. Bull. 64: 856 (2019) [16]; Gild *et al.*, J. Materiomics 5: 337 (2019) [7]; Shivakumar *et al.*, Scripta Mater. 212, 114557 (2022) [12]; Jiang *et al.*, Scripta Mater. 142: 116 (2018) [6]; Gild *et al.*, J. Eu. Ceram. Soc. 38: 3578 (2020) [8]; Wright *et al.*, Acta Mater. 235: 118056 (2022) [23]; Gild *et al.*, Scientific Reports 6: 37946 (2016) [5]; Qin *et al.*, Scripta Mater. 189: 101 (2020) [9]; Qin *et al.*, J. Adv. Ceram. 10: 166 (2021) [10]; Qin *et al.*, J. Eu. Ceram. Soc. 41, 2968 (2021) [11]; and Qin *et al.*, J. Adv. Ceram. 11: 641 (2022) [18].

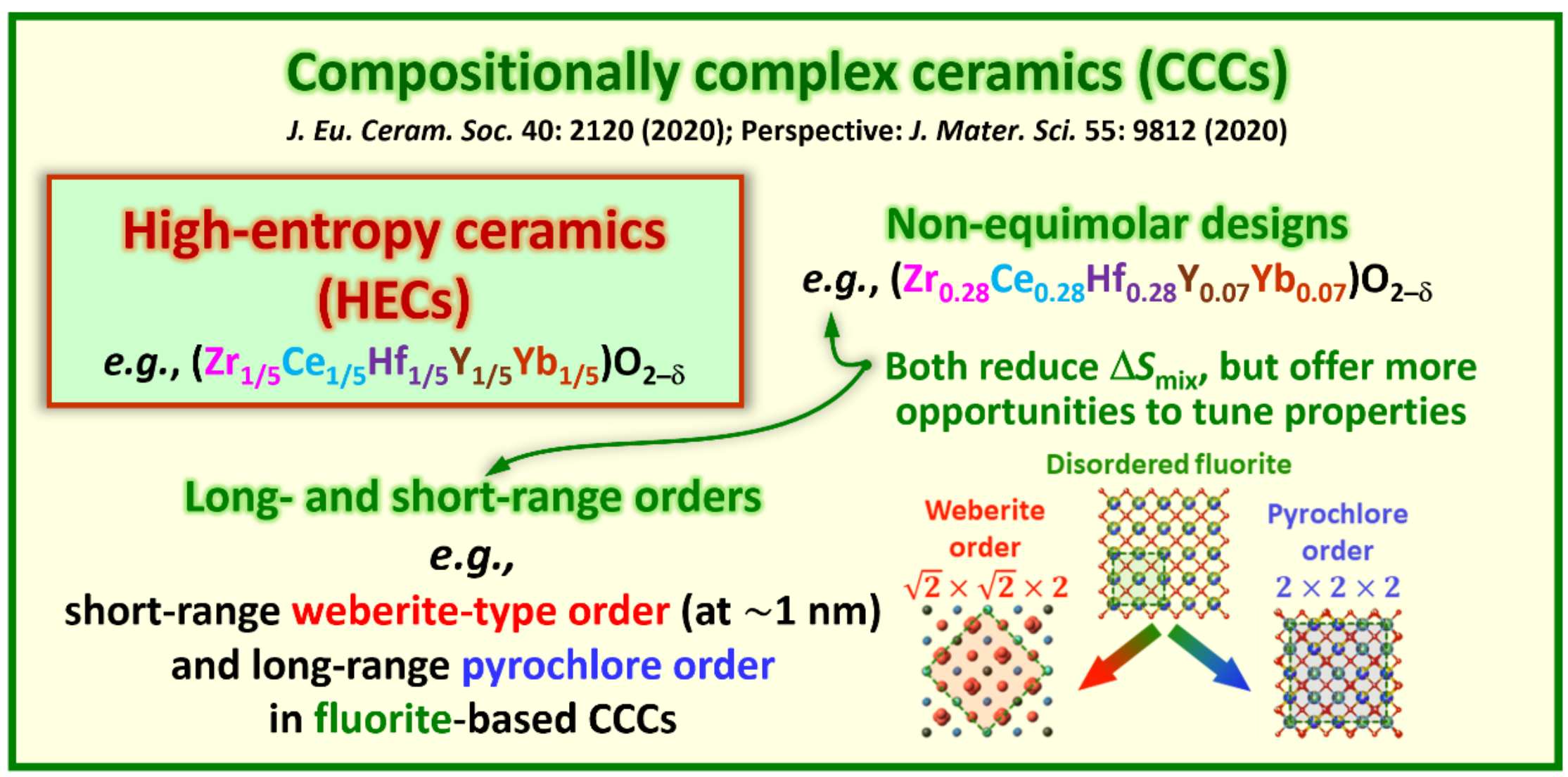

**Fig. 2. From high-entropy ceramics (HECs) to compositionally complex ceramics (CCCs).** CCCs can be defined as ceramic solid solutions with at least three principal components, with HECs (having >1.5 $k_B$ per cation ideal configurational entropy in at least one cation sublattice) largely included as a subset. In a 2020 study [20] and a subsequent perspective article [17], we proposed broadening HECs to CCCs, where non-equimolar compositions and long- or short-range order provide additional opportunities to tune and enhance properties, despite reduced configurational entropy. Fluorite-based HECs and CCCs illustrate these concepts: for example, a non-equimolar composition $(Hf_{0.28}Zr_{0.28}Ce_{0.28}Y_{0.07}Yb_{0.07})_xO_{2-\delta}$ can achieve a superior thermal conductivity-to-modulus ($k/E$) ratio for thermal barrier coating (TBC) applications, outperforming the higher-entropy equimolar $(Hf_{1/5}Zr_{1/5}Ce_{1/5}Y_{1/5}Yb_{1/5})_xO_{2-\delta}$ [20]. Furthermore, weberite-type structural short-range order can exist in long-range disordered fluorite CCCs to achieve ultralow thermal conductivity [23], despite the reduced entropy from such correlated disorder (*i.e.*, a local symmetry lower than the global cubic symmetry of the crystal [24]).

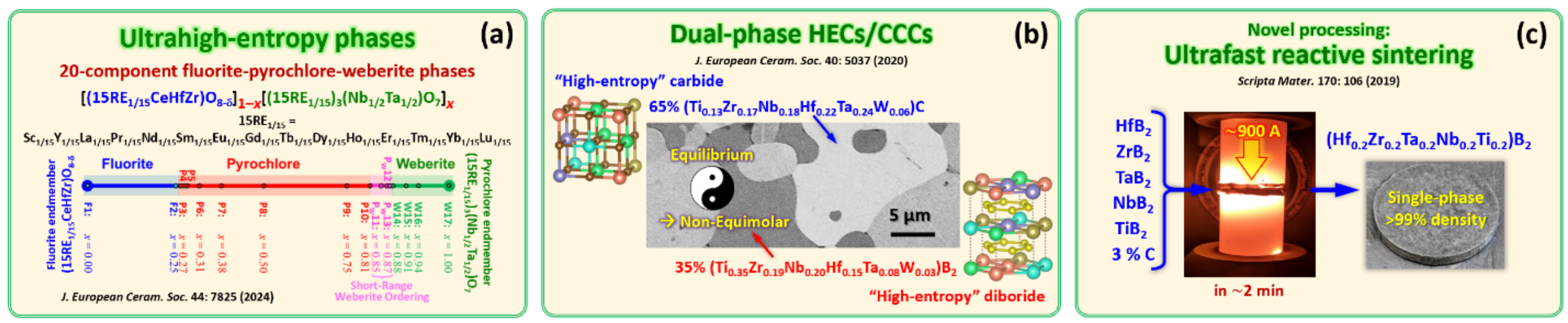

**Fig. 3. Three selected emerging research areas: (a)** ultrahigh-entropy phases exhibiting abrupt phase transitions [19]; **(b)** dual-phase HECs, which transform into two non-equimolar CCC phases at thermodynamic equilibrium despite the overall equimolar composition of the five metal elements [30]; and **(c)** novel synthesis and processing approaches, exemplified by ultrafast reactive sintering [31]. Schematics and images were adapted from Song *et al.*, J. European Ceram. Soc. 44: 7825 (2024) [19], Qin *et al.*, J. European Ceram. Soc. 40: 5037 (2020) [30]; and Gild *et al.*, Scripta Mater. 170: 106 (2019) [31].

## References:

[1] J.W. Yeh, S.K. Chen, S.J. Lin, J.Y. Gan, T.S. Chin, T.T. Shun, et al. Nanostructured high-entropy alloys with multiple principal elements: Novel alloy design concepts and outcomes, Adv Eng Mater 6(5) (2004) 299–303.

[2] B. Cantor, I. Chang, P. Knight, A. Vincent. Microstructural development in equiatomic multicomponent alloys, Mater Sci Eng A 375 (2004) 213–218.

[3] D.B. Miracle, O.N. Senkov. A critical review of high entropy alloys and related concepts, Acta Mater 122 (2017) 448–511.

[4] C.M. Rost, E. Sachet, T. Borman, A. Moballegh, E.C. Dickey, D. Hou, et al. Entropy-stabilized oxides, Nat Commun 6 (2015) 8485.

[5] J. Gild, Y. Zhang, T. Harrington, S. Jiang, T. Hu, M.C. Quinn, et al. High-Entropy Metal Diborides: A New Class of High-Entropy Materials and a New Type of Ultrahigh Temperature Ceramics, Sci Rep 6 (2016) 37946.

[6] S. Jiang, T. Hu, J. Gild, N. Zhou, J. Nie, M. Qin, et al. A new class of high-entropy perovskite oxides, Scr Mater 142 (2018) 116–120.

[7] J. Gild, J. Braun, K. Kaufmann, E. Marin, T. Harrington, P. Hopkins, et al. A high-entropy silicide: $(Mo_{0.2}Nb_{0.2}Ta_{0.2}Ti_{0.2}W_{0.2})Si_2$, J Materiomics 5(3) (2019) 337–343.

[8] J. Gild, M. Samiee, J.L. Braun, T. Harrington, H. Vega, P.E. Hopkins, et al. High-entropy fluorite oxides, J Eur Ceram Soc 38(10) (2018) 3578–3584.

[9] M. Qin, Q. Yan, H. Wang, C. Hu, K.S. Vecchio, J. Luo. High-entropy monoborides: Towards superhard materials, Scr Mater 189 (2020) 101–105.

[10] M. Qin, Q. Yan, Y. Liu, J. Luo. A new class of high-entropy $M_3B_4$ borides, J Adv Ceram 10(1) (2021) 166–172.

[11] M. Qin, Q. Yan, H. Wang, K.S. Vecchio, J. Luo. High-entropy rare earth tetraborides, J Eur Ceram Soc 41(4) (2021) 2968–2973.

[12] S. Shivakumar, M. Qin, D. Zhang, C. Hu, Q. Yan, J. Luo. A new type of compositionally complex M5Si3 silicides: Cation ordering and unexpected phase stability, Scr Mater 212 (2022) 114557.

[13] X. Yan, L. Constantin, Y. Lu, J.-F. Silvain, M. Nastasi, B. Cui. $(Hf_{0.2}Zr_{0.2}Ta_{0.2}Nb_{0.2}Ti_{0.2})C$ high-entropy ceramics with low thermal conductivity, J Am Ceram Soc 101(10) (2018) 4486–4491.

[14] E. Castle, T. Csanádi, S. Grasso, J. Dusza, M. Reece. Processing and Properties of High-Entropy Ultra-High Temperature Carbides, Sci Rep 8(1) (2018) 8609.

[15] T.J. Harrington, J. Gild, P. Sarker, C. Toher, C.M. Rost, O.F. Dippo,et al. Phase stability and mechanical properties of novel high entropy transition metal carbides, Acta Mater 166 (2019) 271–280.

[16] N. Zhou, S. Jiang, T. Huang, M. Qin, T. Hu, J. Luo. Single-phase high-entropy intermetallic compounds (HEICs): bridging high-entropy alloys and ceramics, Sci Bull 64(12) (2019) 856–864.

[17] A.J. Wright, J. Luo. A step forward from high-entropy ceramics to compositionally complex ceramics: a new perspective, J Mater Sci 55(23) (2020) 9812–9827.

[18] M. Qin, H. Vega, D. Zhang, S. Adapa, A.J. Wright, R. Chen, et al. 21-Component compositionally complex ceramics: Discovery of ultrahigh-entropy weberite and fergusonite phases and a pyrochlore-weberite transition, J Adv Ceram 11(4) (2022) 641–655.

[19] K. Song, D. Zhang, K.M. Chung, R. Chen, J. Luo. Fluorite-pyrochlore-weberite phase transitions in a series of 20-component ultrahigh-entropy compositionally complex ceramics, J Eur Ceram Soc 44(13) (2024) 7825–7836.

[20] A.J. Wright, Q. Wang, C. Huang, A. Nieto, R. Chen, J. Luo. From high-entropy ceramics to compositionally-complex ceramics: A case study of fluorite oxides, J Eur Ceram Soc 40(5) (2020) 2120–2129.

[21] A.J. Wright, Q. Wang, S.-T. Ko, K.M. Chung, R. Chen, J. Luo. Size disorder as a descriptor for predicting reduced thermal conductivity in medium- and high-entropy pyrochlore oxides, Scr Mater 181 (2020) 76–81.

[22] F. Gu, W. Wang, H. Meng, Y. Liu, L. Zhuang, H. Yu, et al. Lattice distortion boosted exceptional electromagnetic wave absorption in high-entropy diborides, Matter 8(3) (2025).

[23] A.J. Wright, Q. Wang, Y.-T. Yeh, D. Zhang, M. Everett, J. Neuefeind, et al. Short-range order and origin of the low thermal conductivity in compositionally complex rare-earth niobates and tantalates, Acta Mater 235 (2022) 118056.

[24] A. Simonov, A.L. Goodwin. Designing disorder into crystalline materials, Nat Rev Chem 4(12) (2020) 657–673.

[25] D. Zhang, Y. Chen, H. Vega, T. Feng, D. Yu, M. Everett, et al. Long- and short-range orders in 10-component compositionally complex ceramics, Adv Powder Mater 2(2) (2023) 100098.

[26] M. Qin, J. Gild, H. Wang, T. Harrington, K.S. Vecchio, J. Luo. Dissolving and stabilizing soft $WB_2$ and $MoB_2$ phases into high-entropy borides via boron-metals reactive sintering to attain higher hardness, J Eur Ceram Soc 40(12) (2020) 4348–4353.

[27] L. Feng, W.G. Fahrenholtz, G.E. Hilmas, Y. Zhou, J. Bai. Strength retention of single-phase high-entropy diboride ceramics up to 2000° C, J Am Ceram Soc 107(3) (2024) 1895–1904.

[28] M. Zhang, S. Lan, B.B. Yang, H. Pan, Y.Q. Liu, Q.H. Zhang, et al. Ultrahigh energy storage in high-entropy ceramic capacitors with polymorphic relaxor phase, Science 384(6692) (2024) 185–189.

[29] A. Navrotsky. Thermodynamics of $A_3O_4$-$B_3O_4$ spinel solid solutions, J Inorg Nucl Chem 31(1) (1969) 59–72.

[30] M. Qin, J. Gild, C. Hu, H. Wang, M.S.B. Hoque, J.L. Braun, et al. Dual-phase high-entropy ultra-high temperature ceramics, J Eur Ceram Soc 40(15) (2020) 5037–5050.

[31] J. Gild, K. Kaufmann, K. Vecchio, J. Luo. Reactive flash spark plasma sintering of high-entropy ultrahigh temperature ceramics, Scr Mater 170 (2019) 106–110.

[32] H. Xie, M. Qin, M. Hong, J. Rao, M. Guo, J. Luo, et al. Rapid liquid phass-assisted ultrahigh-temperature sintering of high-entropy ceramic composites, Sci Adv 8(27) (2022) eabn8241.

[33] C. Wang, W. Ping, Q. Bai, H. Cui, R. Hensleigh, R. Wang, et al. A general method to synthesize and sinter bulk ceramics in seconds, Science 368(6490) (2020) 521–526.

[34] Z. Wen, Z. Tang, Y. Liu, L. Zhuang, H. Yu, Y. Chu. Ultrastrong and High Thermal Insulating Porous High-Entropy Ceramics up to 2000 °C, Adv Mater 36(14) (2024) 2311870.

[35] J. Luo. Computing grain boundary “phase” diagrams, Interdiscip Mater 2(1) (2023) 137-160.

[36] S.-T. Ko, T. Lee, J. Qi, D. Zhang, W.-T. Peng, X. Wang, et al. Compositionally complex perovskite oxides: Discovering a new class of solid electrolytes with interface-enabled conductivity improvements, Matter 6(7) (2023) 2395–2418.

[37] S.-T. Ko, C. Du, H. Guo, H. Vahidi, J.L. Wardini, T. Lee, et al. Temperature-dependent microstructural evolution in a compositionally complex solid electrolyte: The role of a grain boundary transition, J Adv Ceram 14(3) (2025) 9221047.

[38] J. Luo, N. Zhou. High-entropy grain boundaries, Commun Mater 4(1) (2023) 7.

[39] K.M. Chung, S.R. Adapa, Y. Pei, R.H. Yeerella, L. Chen, S. Shivakumar, et al. Low Thermal Conductivity and Diffusivity at High Temperatures Using Stable High-Entropy Spinel Oxide Nanoparticles, Adv Mater 37(6) (2025) 2406732.